\documentclass[aps,prb,twocolumn,multicolumn,graphics,showpacs]{revtex4}
\input epsf
\input{epsf.sty}

\begin{document}

\hsize\textwidth\columnwidth\hsize
\csname@twocolumnfalse\endcsname

\title{ Vortex Solid-Liquid Transition in
Bi$_{2}$Sr$_{2}$CaCu$_{2}$O$_{8+\delta}$ with a High Density of Strong Pins}

\author {S. Colson, C.J. van der Beek, M. Konczykowski}
\affiliation{\mbox{Laboratoire des Solides Irradi\'{e}s, CNRS-UMR 7642 
\& CEA/DSM/DRECAM, Ecole Polytechnique, 91128  Palaiseau, France}}

 \author {M.B. Gaifullin, Y. Matsuda}
\affiliation{Institute for Solid State Physics, University     
of Tokyo, Kashiwanoha, Kashiwa, Chiba 277-8581, Japan}

\author{P. Gier\l owski}
\affiliation{Institute of Physics, Polish Academy of           
Sciences, Al. Lotnikow 32/46, 02-668 Warsaw, Poland}

\author{Ming Li, P.H. Kes}
\affiliation{\mbox{Kamerlingh Onnes Laboratorium, Leiden             
University, P.O. Box 9506,  2300 RA Leiden, The Netherlands}}

\date{\today}

\begin{abstract}
\noindent
The introduction of a large density of columnar defects in 
Bi$_{2}$Sr$_{2}$CaCu$_{2}$O$_{8+\delta}$ crystals does not, at sufficiently low 
vortex densities, increase the irreversibility line beyond the first order 
transition (FOT) field of pristine crystals. At such low fields, the flux line 
wandering length $r_{w}$ behaves as in pristine 
crystals. Next, vortex positional correlations along the $c$--axis in the vortex
Bose glass at fields above the FOT are smaller than in the low--field vortex 
solid. Third, the Bose-glass-to-vortex liquid transition is signaled by a rapid 
decrease in $c$-axis phase correlations. These observations are 
understood in terms of the ``discrete superconductor'' model.

\end{abstract}
\pacs{74.25.Op,74.25.Qt,74.25.Nf}

\maketitle                             
%

Heavy-ion irradiation of the layered high temperature superconductor 
Bi$_{2}$Sr$_{2}$CaCu$_{2}$O$_{8+\delta}$ induces the formation of 
amorphous latent tracks in the material, that act as strong pinning centers 
for vortices. Vortex localization on these ``columnar defects'' (CD's) 
leads to the superconducting Bose-glass phase at low field and 
temperature.~\cite{Nelson92} In single crystalline
Bi$_{2}$Sr$_{2}$CaCu$_{2}$O$_{8+\delta}$, the irreversibility line 
(IRL) $B_{irr}(T)$ [or $T_{irr}(B)$], below which the $I(V)$ characteristic is no longer linear
because of vortex pinning by CD's, can be identified with the Bose-glass 
transition line.~\cite{Beek01} It was proposed\cite{Beek01} that when 
the defect density $n_{d}$ is much larger than the vortex density $B/\Phi_{0}$ 
(with $\Phi_{0} = h/2e$ the flux quantum), vortex lines will behave qualitatively 
as in unirradiated Bi$_{2}$Sr$_{2}$CaCu$_{2}$O$_{8+\delta}$ crystals, even though 
the two-dimensional (2D) pancake vortices of which they are 
constituted are always located on the tracks. 
This ``discrete superconductor'' description\cite{Beek01} can be applied  
when $B, B_{\Lambda} < \frac{1}{6} B_{\phi}$, where $B_{\phi} = 
\Phi_{0}n_{d}$ and $B_{\Lambda} = \Phi_{0}( \lambda_{J}^{-1} + 
\lambda_{ab}^{-1})^{2}$.\cite{Beek01} 
 The Josephson length $\Lambda_{J} = \gamma s$,
$\gamma = \lambda_{c} / \lambda_{ab}$ is the ratio of the London penetration depths 
$\lambda_{ab,c}$ for currents parallel to the material $ab$ plane and 
$c$ axis respectively, and  $s = 1.5$ nm is the separation between 
superconducting layers.


Here, we present measurements of vortex fluctuations in heavy-ion 
irradiated (HII) Bi$_{2}$Sr$_{2}$CaCu$_{2}$O$_{8+\delta}$ crystals, near the IRL,
that corroborate the ``discrete superconductor'' model. The first 
order vortex solid to liquid transition (FOT) in pristine 
Bi$_{2}$Sr$_{2}$CaCu$_{2}$O$_{8+\delta}$ crystals~\cite{Zeldov95} has 
a very different dependence on oxygen content $\delta$ than the second order
Bose glass to flux liquid transition after HII. The modification of 
the relative position of the two transitions in the vortex phase diagram
by oxygen doping exposes two regimes. At high temperatures and low fields, the vortex 
solid-to-liquid transition in HII Bi$_{2}$Sr$_{2}$CaCu$_{2}$O$_{8+\delta}$ coincides 
with the FOT before irradiation. Moreover, the vortex wandering length 
$r_{w}$ obeys the same temperature and field dependence as in the pristine 
material,~\cite{Colson03} suggesting an identical mechanism of the transition. 
At higher vortex densities and lower temperature, vortex lines in the 
Bose glass are surprisingly {\em less} correlated along the c-axis 
than in the vortex solid before HII. Nevertheless, the Bose glass-to-flux liquid 
transition is also marked by a rapid decrease of $c$--axis correlations.

We have used underdoped ($T_{c} = 69.4\pm 0.6$~K) 
Bi$_{2}$Sr$_{2}$CaCu$_{2}$O$_{8+\delta}$ single crystals, grown by the travelling 
solvent floating zone method at the FOM-ALMOS center, the Netherlands, in 25~mbar 
O$_{2}$ partial pressure.~\cite{MingLi02}  The crystals were annealed 
for one week in flowing N$_{2}$ gas, and irradiated with 5.8~GeV 
Pb$^{56+}$ ions at GANIL (Caen, France). Crystal A 
(dimensions $610 \times 420 \times 30$ $\mu$m$^{3}$, $B_{\phi} = 2$ T)
was irradiated at $T = 80$ K to avoid self-doping;\cite{Li02} indeed, $T_{c}$
was unchanged following irradiation. The $T_{c}$ of crystals B and 
C (dimensions $980 \times 600 \times 40$ and $1160 \times 660 \times 40$ 
$\mu$m$^{3}$, $B_{\phi}=1$~T), irradiated at 293 K, increased to 75.6 $\pm$0.3~K.\cite{Li02}
Crystals A and C were mounted with the $c$ axis parallel to the ion beam during
irradiation, whereas sample B was rocked at incommensurate 
frequencies, around two orthogonal axes, resulting in homogeneously splayed
CD's with angles between 0 and 15 degrees with respect to the $c$~axis. 

\begin{figure}[t!]  
 	\centerline{\epsfxsize 8.5cm \epsfbox{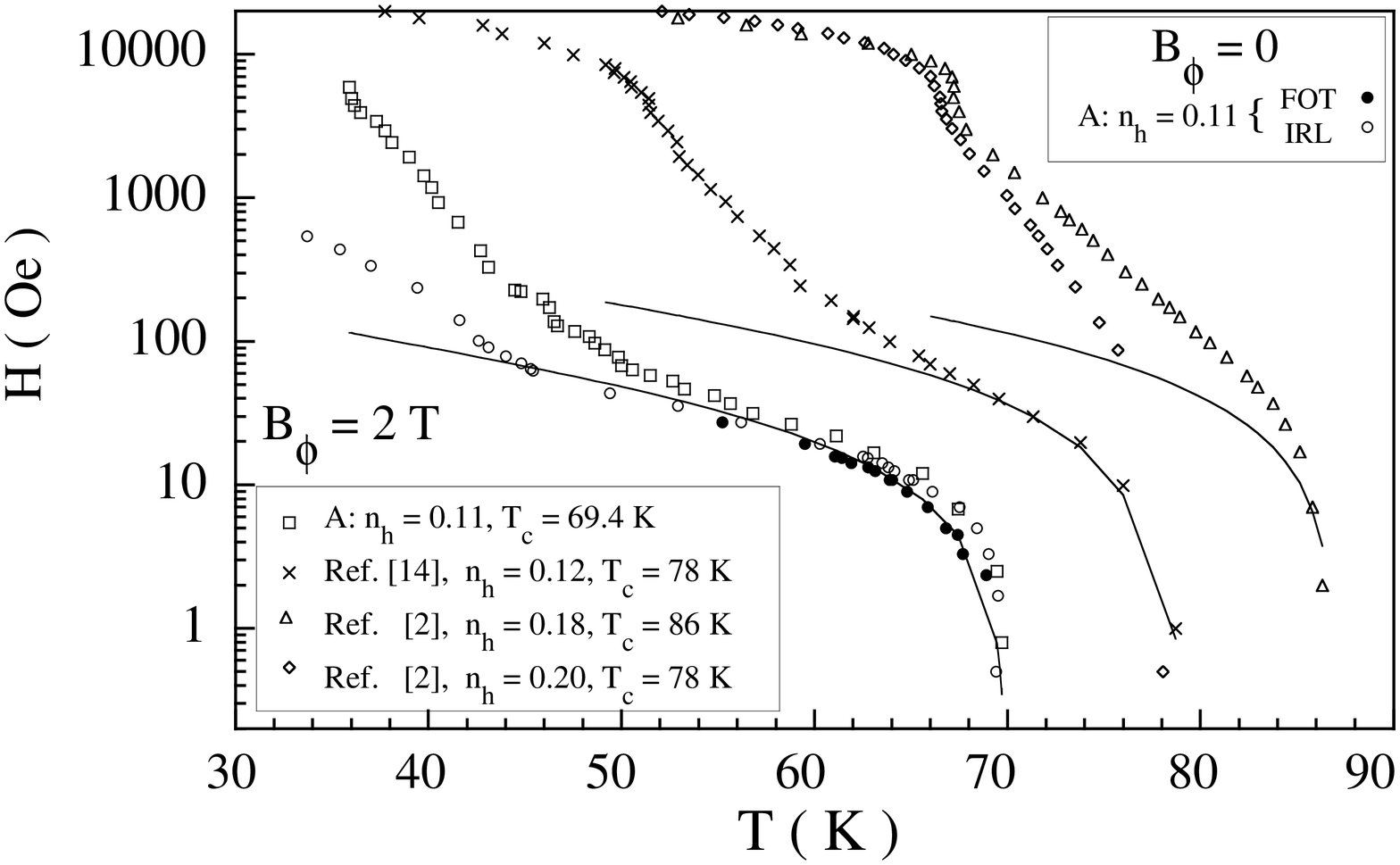}} 
	\caption{IRL (open symbols) and FOT lines (closed symbols, drawn lines) in 
	Bi$_{2}$Sr$_{2}$CaCu$_{2}$O$_{8+\delta}$ single crystals with different oxygen 
	content, before and after irradiation with $1 \times10^{11} {\rm cm^{-2}}$ 
	6 GeV Pb$^{56+}$ ions ($B_{\phi} = 2$ T). ($\bullet$, $\circ$) 
	pristine underdoped crystal A, $n_{h} = 0.11, T_{c} = 69.4$ K; 
	(\protect\raisebox{1mm}{\protect\framebox{}}) 
	crystal A, two days after irradiation at 80 K. 
	($\times$, $\triangle$, $\diamond$) different doping levels $n_{h}$, 
	from Refs.~\protect\onlinecite{Beek01} and \protect\onlinecite{Li02-Crete}.
	The FOT lines of the pristine material (drawn lines) are 
	well described by Eq.~(\protect\ref{eq:FOT}), with parameter values 
	($\varepsilon_{0}s/k_{B}, B_{\Lambda}$)  = (360 K, 35 G) ($n_{h} = 
	0.11$), (475 K, 55 G) ($n_{h} = 0.12$), (1000 K, 45 G) ($n_{h}=0.18$). }
	\label{Fig1}
\end{figure}

The IRL's of different crystals (Fig.~\ref{Fig1}) were obtained as the onset temperature of the nonlinear ac 
transmittivity using a local Hall probe magnetometer in ac mode, 
operated at a frequency of 21~Hz and an ac field amplitude of 0.5~Oe.~\cite{Gilchrist93,Beek95} 
The differential magneto-optical technique (DMO)~\cite{Soibel00,Soibel2001,Beek2003} was
used to determine the IRLs of samples A, B and C (inset of Fig.~\ref{Fig4}a).  
DMO reveals that $B_{irr}$ depends on the position on the crystal surface;
\cite{Soibel2001,Beek2003} the range of $B_{irr}$ is denoted by the error bars. 

Figure~\ref{Fig1} shows the dependence of $B_{irr}(T)$ after HII for  
Bi$_{2}$Sr$_{2}$CaCu$_{2}$O$_{8+\delta}$ crystals with different 
oxygen content. The oxygenation can be characterized by the 
number of holes $n_{h}$ per Cu.\cite{Li02,Presland} For the 
as--grown ($n_{h} = 0.18$) and overdoped ($n_{h} = 0.2$)
crystals of Ref.~\onlinecite{Beek01}, that have a transition width $\Delta T_{c}= 
0.3$ K, the IRL coincides with the FOT of pristine 
Bi$_{2}$Sr$_{2}$CaCu$_{2}$O$_{8+\delta}$ with the same $T_{c}$ over a temperature 
range of less than 1 K below $T_{c}$; for lower temperature, one observes 
the usual dramatic increase of $B_{irr}$ following HII.~\cite{Konczykowski95} 
For moderate underdoping, $n_{h} = 0.12$,\cite{Li02-Crete} the 
IRL's prior to and after irradiation coincide over a 
span of nearly 10 K. Finally, for crystal A (as well as for crystals B and C, 
not shown in Fig.1) an increase of 
$B_{irr}$ after irradiation is observed only for $T < 57 \pm 2$~K. In the 
temperature regime in which an irradiation-induced enhancement  
is observed, $B_{irr}$  follows an exponential temperature 
dependence for all doping levels. The IRL always lies below the upper limit 
\cite{Beek01,Li02-Crete} 
\begin{equation}
	B_{irr}^{max}(T) = B_{\Lambda}
	\left(\frac{\varepsilon_{0}(T)s}{k_{B}T}\right)
	\exp\left(\frac{\varepsilon_{0}(T)s}{k_{B}T}\right)
	   \hspace{3mm}
	(B, B_{\Lambda} \ll B_{\phi} ).
	\label{eq:T-dep-Birr}
\end{equation}
\noindent This represents a ``mobility threshold'' or 
``delocalization line'' above which two-dimensional pancake vortices can 
diffuse from their equilibrium site in the vortex solid. The IRL can 
be compared to the FOT line of the pristine crystal. The latter is well decribed by 
the expression\cite{Glazman91} 
\begin{equation}
    B_{FOT} = 0.5 B_{\Lambda}  \varepsilon_{0}(T)s/k_{B}T,
    \label{eq:FOT}
\end{equation}
drawn in Fig. 1 using values of $\lambda_{ab}$ and $\gamma$
determined from independent experiments.\cite{Colson03,Li02} Note that $B_{irr}^{max}(T)$ 
and the FOT line depend only on the vortex line energy per unit 
length, $\varepsilon_{0}(T)=\Phi_{0}^{2}/4\pi 
\mu_{0}\lambda_{ab}^{2}(T)$, and on $B_{\Lambda}$.

The evolution of the vortex phase diagram with changing oxygen content  
results from the different dependence of the Bose-glass 
delocalization line and the FOT line on $\varepsilon_{0}(T)$.
As one goes from optimally doped to underdoped 
Bi$_{2}$Sr$_{2}$CaCu$_{2}$O$_{8}$, the decrease of $\varepsilon_{0}(T)$
leads to a downward shift of the exponential Bose-glass line 
that is much larger than the roughly linear shift of the FOT, \cite{Koshelev97}
uncovering a substantial temperature range over which 
the FOT line survives the irradiation (Fig.~1). In optimally doped and 
overdoped crystals the relative shift is much smaller, but, 
nevertheless, one can identify a temperature range near $T_{c}$ in which HII does \em not \rm
increase the IRL beyond the FOT line of the pristine material.

\begin{figure}[b!]
		\centerline{\epsfxsize 8cm \epsfbox{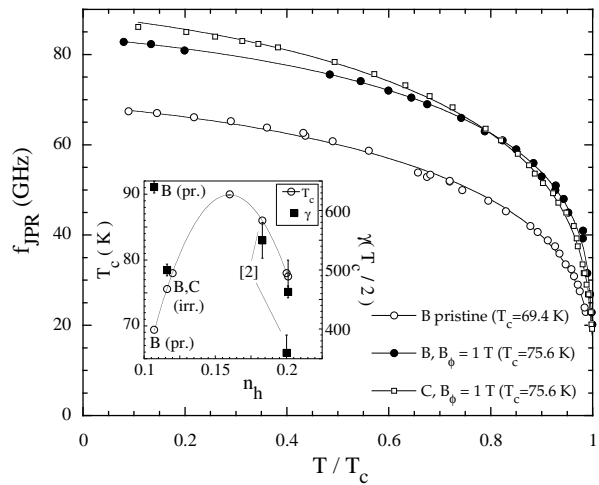}}
		\caption{Josephson plasma frequency $f_{JPR}$ in zero field
		vs $T/T_{c}$ for crystal B before irradiation, and for crystals B
		and C after irradiation. The increase in $f_{JPR}$ is due to 
		oxygen self-doping.\cite{Li02} Solid lines are guides to the eye.
		Inset: $T_{c}$ and $\gamma$ at $\frac{1}{2}T_{c}$ vs. the hole
		density per Cu $n_{h}$, for pristine (pr.) and irradiated crystals B 
		and C, and for the crystals of Ref.~\protect\onlinecite{Beek01}. }
		\label{Fig2}
\end{figure}


In order to characterize vortex fluctuations in this high temperature 
regime, we have determined the vortex wandering length in the vortex 
solid\cite{Bulaevskii00,Koshelev00} in the presence of CD's using Josephson
Plasma Resonance (JPR).\cite{Colson03,Matsuda95} In Bi$_{2}$Sr$_{2}$CaCu$_{2}$O$_{8+\delta}$, 
thermal fluctuations may shift two pancake vortices belonging to the same flux
line, and located in consecutive superconducting layers $n$ and
$n+1$, by a vector ${\bf u}_{n,n+1}$ with respect to one another. 
The wandering length $r_{w}$ is defined as the thermal and disorder
average $r_{w}=\langle {\bf u}_{n,n+1} \rangle$, and can be extracted 
from JPR experiments. Following the procedure of 
Ref.~\onlinecite{Colson03}, valid in the regime $B < B_{\Lambda} \sim 30-40$ G,
\begin{equation}
	r_{w}^{2}  = \frac {2 \Phi_{0}}{\pi B} 
	\left[1-\frac{f_{JPR}^{2}(B,T)}{f_{JPR}^{2}(0,T)}\right]
 ,
	\label{eq:rw-Sylvain}
\end{equation}
\noindent with $f_{JPR}(B,T)$ and $f_{JPR}(0,T)$ the JPR
frequencies in field $B$ and in zero field, respectively. The ratio 
$f_{JPR}^{2}(B,T) / f_{JPR}^{2}(0,T) \equiv \langle \cos(\phi_{n,n+1}) \rangle$
corresponds to the average cosine of the gauge-invariant difference 
of the superconducting order parameter phase between layers 
$n$ and $n+1$.~\cite{Bulaevskii95}

\begin{figure}[t!] 
		\centerline{\epsfxsize 7.5cm \epsfbox{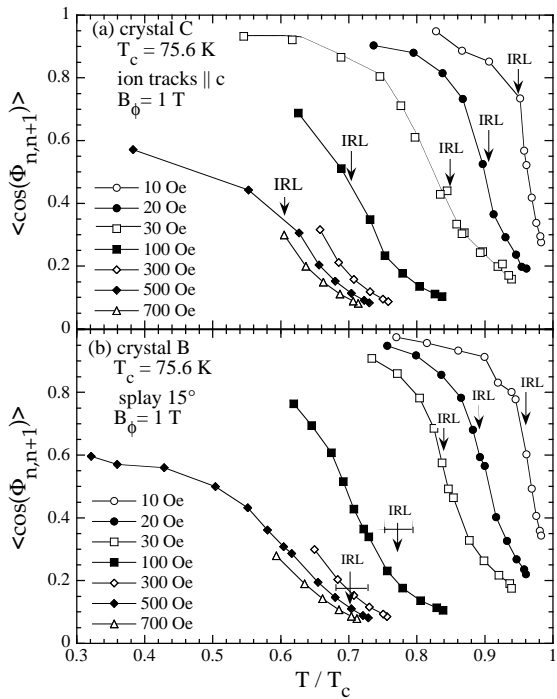}}
		\caption{ $ \langle \cos \phi_{n,n+1} \rangle$ vs. reduced
		temperature in irradiated samples C (a) and B (b), 
		for different fields 10 Oe$< H <$ 700 Oe. Arrows show  
		$T_{irr}$ obtained by DMO. }
		\label{Fig3}
\end{figure}

The JPR frequency in zero field was measured using the bolometric 
method~\cite{Matsuda94} on crystal~B both before and after irradiation, and on
crystal C after irradiation (Fig.~\ref{Fig2}). We observe an increase in 
$f_{JPR}$ after room temperature irradiation, implying a decrease in the London 
penetration depth  $\lambda_{c} = 2\pi/\sqrt{\mu_{0}\epsilon} f_{JPR}(0,T)$. 
Here $\epsilon=11.5 \epsilon_{0}$ is the dielectric permittivity, 
$\epsilon_{0}$ is the permittivity of the vacuum,~\cite{Gaifullin01} 
and $\mu_{0} = 4 \pi \times 10^{-7}$ Hm$^{-1}$. The decrease of 
$\lambda_{c}$ can be understood as stemming from the oxygen self-doping
induced by the irradiation,\cite{Li02} if one assumes that oxygen
ions are expelled from the tracks not only to the CuO$_{2}$ planes
but also to the BiO planes. This would lead to a decrease of the $c$ 
axis parameter,~\cite{Li96} an increase of the interlayer coupling, 
and a concommitant decrease of $\lambda_{c}$. 

\begin{figure} [t!] 
\centerline{\epsfxsize 7.9cm \epsfbox{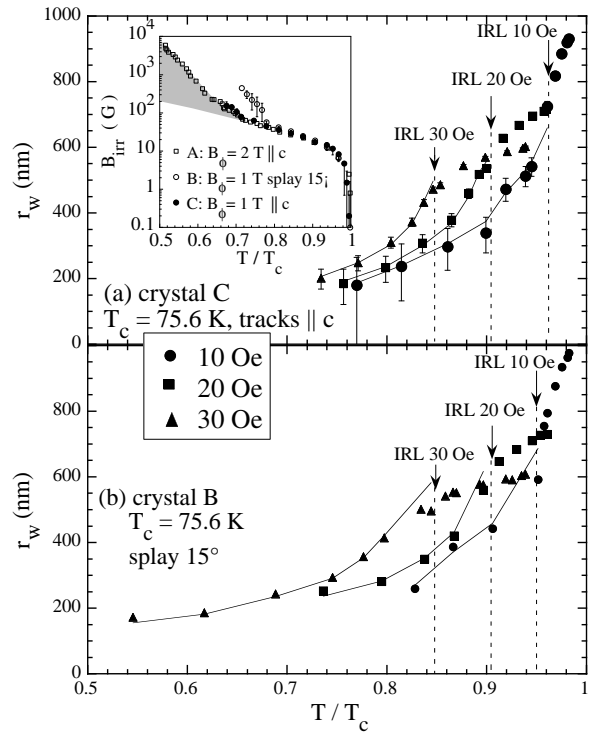}}
\caption{ Experimental $r_{w}$ vs $T/T_{c}$ for irradiated
	  samples B and C, for low field values~: $H=10$ Oe, 20 Oe and 30 Oe. Solid
	  lines are fits to Eq.~(\ref{eq:rw-T-Bdependence}), while dashed lines 
	  denote $T_{irr}$ for the different fields. The $(B_{irr},T)$--diagram
	  for crystals A, B (from DMO, see text) and C is shown in the inset.}
\label{Fig4}
\end{figure}

Both the bolometric method and the cavity perturbation
technique~\cite{Colson03} were used to measure $\langle
\cos(\phi_{n,n+1})\rangle$ on crystals~B and~C, in fields 
up to 700~Oe (Fig.~\ref{Fig3}). For $B \leq 30$~G, the IRL roughly coincides 
with the inflection point of  $\langle \cos(\phi_{n,n+1})\rangle$ 
as function of $T$, as does the FOT in unirradiated samples.~\cite{Gaifullin00,Colson03} 
From $\langle \cos(\phi_{n,n+1})\rangle$, we obtain $r_{w}$ using 
Eq.~(\ref{eq:rw-Sylvain}).
The field and temperature dependence of $r_{w}$, plotted in
Fig.~\ref{Fig4}, is similar to that observed in pristine
Bi$_{2}$Sr$_{2}$CaCu$_{2}$O$_{8+\delta}$,~\cite{Colson03} at least 
for $T<T_{irr}(B)$. Thus, it is tempting to fit the
experimental $r_{w}$ to the formula that describes
vortex fluctuations in the {\em vortex solid} in pristine crystals,\cite{Colson03}
\begin{eqnarray}
	r_{w}^{2} & \approx  & \alpha s^{2}
	\frac{k_{B}T\gamma^{2}}{\varepsilon_{0}s}
			       \left[   \frac{4}{ \pi \left(\alpha x^{2}  +
					 \frac{1}{4} \right) }  +
					 \frac{1}{2} \ln\left(0.66 x^{2} \right) \right.
	\nonumber \\
	 & &     \hspace{1cm}       \left. + \frac{2}{\pi^{2}} \left(
			       \frac{a_{0}}{\lambda_{ab}} \right)^{2}
				    \ln \left(1+\frac{x^{2}}{21.3} \right)
				  \right]^{-1}.
	\label{eq:rw-T-Bdependence}
\end{eqnarray}
\noindent Here, $x=a_{0}/r_{w}$, $a_{0}\approx (\Phi_{0}/B)^{1/2}$ is the
intervortex spacing, and $\alpha \sim 1$; $\lambda_{ab}(0)=358$~nm is 
obtained from $\lambda_{ab}$ of pristine underdoped Bi$_{2}$Sr$_{2}$CaCu$_{2}$O$_{8+\delta}$
($T_{c}=69.4$~K) \cite{Colson03} and the Uemura relation,~\cite{Uemura89} and 
$\lambda_{c}$ from $f_{JPR}(0)$.  Using Eq.~(\ref{eq:rw-T-Bdependence}), 
we obtain a very good description of $r_{w}$ in the regime $T<T_{irr}(B)$ 
(Fig.~\ref{Fig4}). As in unirradiated Bi$_{2}$Sr$_{2}$CaCu$_{2}$O$_{8+\delta}$
crystals, the whole expression (\ref{eq:rw-T-Bdependence})
is used to fit the data at the lowest field $H=10$~Oe (with $\alpha = 0.6-0.75$).
At $H=20$ and 30~Oe, Eq.~(\ref{eq:rw-T-Bdependence}) also well describes 
the data, but, as in pristine Bi$_{2}$Sr$_{2}$CaCu$_{2}$O$_{8+\delta}$,  
better fits (with $\alpha =0.35-0.5$) are obtained if the first term 
in the denominator is omitted. We have shown previously that pancake
vortices are always located on the defects,\cite{Li02-Crete,Beek00} 
which outnumber the flux lines by a factor 300 to 1000.
Nevertheless, in the temperature range in which the IRL is described 
by Eq.~(\ref{eq:FOT}), thermal vortex excursions in HII 
Bi$_{2}$Sr$_{2}$CaCu$_{2}$O$_{8+\delta}$ are indistinguishable
from those in the pristine material. A similar 
conclusion may well hold true at lower temperature and higher fields, 
as follows from the recent observation in HII underdoped
Bi$_{2}$Sr$_{2}$CaCu$_{2}$O$_{8+\delta}$ of an abrupt change in vortex 
dynamics {\em below} the exponential Bose-glass line, at the FOT field 
prior to irradiation.\cite{Marcin-Rio} In other words, vortex fluctuations 
in the vortex solid phase are not affected by the introduction of a high 
density of columnar pins.

On the contrary, in the part of the vortex liquid phase that is converted 
to Bose-glass by the irradiation (hatched area in the Inset to 
Fig.~\ref{Fig4}a), vortex fluctuations aare profoundly affected by 
the CD's. In pristine Bi$_{2}$Sr$_{2}$CaCu$_{2}$O$_{8+\delta}$, order parameter
phase correlations in the liquid are described by the high--temperature expansion
result $\langle \cos(\phi_{n,n+1}) \rangle \propto 1 / T B$.
\cite{Koshelev97,Gaifullin00} After irradiation, $\langle \cos(\phi_{n,n+1})\rangle$ 
in this field region no longer follows this behavior, but rather 
follows a convex temperature dependence, saturating
at a low temperature value of approximately 0.6 (Fig.~\ref{Fig3}). 
The low-temperature saturation means that between the FOT and the Bose glass line, vortex lines (pancake stacks) 
are {\em less} ordered along the $c$~axis than in the vortex solid, but {\em more}
ordered than in the flux liquid before irradiation. Presumably, this is because the 
pancakes belonging to the same flux line occupy many different columns even in the 
Bose glass phase.\cite{Beek01,Beek00} Nevertheless, pancakes stay confined to the 
same ``site'', or vortex line, which leads to an enhancement of $c$-axis 
correlations.  A possible mechanism for this enhancement is that the columnar 
defects play the role of a ``substrate potential'', supplementary to 
the electromagnetic and Josephson coupling between 
pancakes.\cite{Dodgson99} It can be seen from Fig.~\ref{Fig3} that 
the $c$-axis phase coherence induced by the CD's is now destroyed at the 
IRL rather than at the FOT: At $T_{irr}$, $\langle \cos(\phi_{n,n+1})\rangle$ 
shows the inflexion point characteristic of the crossover from a state with 
long-range order of the superconducting phase.\cite{Gaifullin00}
This result confirms that of Doyle {\em et al.} that the IRL in HII 
layered superconductors corresponds to a loss of $c$ axis phase coherence.\cite{Doyle96}

To conclude, we have shown that the introduction of a large density 
of amorphous columnar defects does not change thermal vortex 
excursions in the vortex solid state of a layered superconductor. At temperatures
sufficiently close to $T_{c}$, at which the defects are ineffective, this has 
the consequence that the irreversibility line in the presence of CD's 
coincides with that of the pristine superconductor. The situation is 
radically different in the flux liquid: CD's increase pancake 
alignment and phase correlations along the $c$-axis with respect to the 
situation in the flux liquid, but cannot enhance them beyond the 
correlations in the vortex solid.




 
\end{document}